\documentclass[useAMS, usenatbib]{mn2e_letter}

\usepackage{graphicx,deluxetable}
\usepackage{amsmath, amsfonts, amssymb, fancyhdr,times}
\usepackage[usenames,dvipsnames]{color}
\usepackage{natbib}
\usepackage{hyperref}

\newcommand{\lco}{L_{\rm CO}}
\newcommand{\lhcn}{L_{\rm HCN}}
\newcommand{\lfir}{L_{\rm FIR}}
\newcommand{\lir}{L_{\rm IR}}

\newcommand{\cms}{\;{\rm cm}^{-2}}
\newcommand{\thi}{\bar{\tau}_{1667}}
\newcommand{\tlo}{\bar{\tau}_{1665}}

\newcommand{\kms}{\;{\rm km\; s^{-1}}}

\newcommand{\atd}{\rm ATLAS^{3D}}
\newcommand{\hi}{H~{\sc i}}

\title{Observations of hydroxyl in early-type galaxies}
\author[McBride, Alatalo, \& Nyland]{James McBride$^1$\thanks{E-mail: jmcbride@berkeley.edu}, Katherine Alatalo$^2$, and Kristina Nyland$^{3,4}$ \\
$^{1}$Department of Astronomy, Campbell Hall, University of California, Berkeley, California 94720, USA\\
$^{2}$Infrared Processing and Analysis Center, California Institute of Technology, Pasadena, California 91125, USA\\
$^{3}$Netherlands Institute for Radio Astronomy (ASTRON), Postbus 2, 7990 AA Dwingeloo, The Netherlands\\
$^{4}$Physics Department, New Mexico Tech, Socorro, NM 87801, USA
}

\begin{document}
\maketitle
\begin{abstract}
We used Arecibo Observatory and the Green Bank Telescope to observe OH in twelve early-type galaxies with known reservoirs of dense gas. 
We present three new detections of OH in absorption in the 1667~MHz line.  
One objective of our survey was to find evidence of molecular outflows, but our sensitivity, and the strength of the OH absorption, were insufficient to detect outflows.
The detected sources have infrared luminosities and dust temperatures among the lowest of any galaxy detected in OH absorption.  
The ratio $\lhcn / \lco$, a measure of the dense gas fraction in galaxies, is a powerful selector of OH megamasers for galaxies with high infrared luminosity.  
In early-type galaxies, which have much lower infrared luminosities, $\lhcn / \lco$ is also a promising tool for discovering OH, but in absorption rather than in maser emission. 
In addition to dense molecular gas, a radio continuum source and a favorable line-of-sight to the observer are likely key factors in detecting OH absorbers. \\ \\
{\bf Key words:} galaxies: ISM --- galaxies: elliptical and lenticular, cD --- galaxies: evolution

\end{abstract}

\section{Introduction} \label{intro}
The first detections of OH outside the Milky Way were in the starburst galaxies NGC~253 and Messier~82 \citep{Weliachew1971}.  
Both galaxies were originally seen only in absorption, but later observations revealed maser emission as well \citep{Gardner1975,Nguyen1976}. 
In the ultra-luminous infrared galaxy (ULIRG) Arp~220, \citet{Baan1982} discovered OH masing with a luminosity millions of times larger than any masers in the Milky Way. 
Arp~220, and similarly powerful OH maser galaxies, are now called OH megamasers (OHMs).
Subsequent surveys of OH targeted galaxies based on perceived similarity to the starburst galaxies in which OH was first detected.
Criteria for inclusion in early surveys of OH include neutral hydrogen absorption, strong radio continuum, or prominent foreground molecular disks \citep[e.g.,][]{Baan1985a,Schmelz1986}.

As more OHMs and absorbers were discovered, \citet{Baan1989} showed that both populations had optically thick molecular clouds along the line-of-sight to a nuclear radio continuum source.  
The far infrared (FIR) colors of the populations were distinct: masers occurred in galaxies with warm FIR colors, while galaxies with cooler FIR colors absorbed OH. 
\citet{Baan1989} also found that OHM luminosity increased superlinearly with FIR luminosity, prompting surveys for OH targeting LIRGs \citep{Darling2000,Darling2001,Darling2002a}.
\citet{Darling2007} found that in addition to high $\lfir$, OHMs have large dense gas fractions, as measured by $\lhcn / \lco$, and suggested a threshold $\lhcn / \lco > 0.07$ above which OHM emission is triggered.

The recent discovery of reservoirs of molecular gas in early-type galaxies (ETGs), including tracers of dense gas such as HCN, offers an opportunity to study OH in galaxies with properties of $\lfir$, $\lco$, and $\lhcn$ unlike any that have previously been observed.
ETGs are one of two basic types of galaxies, with the other being late-type galaxies.
Late-type galaxies have spiral structure and prominent disks, and are generally blue and star forming. 
ETGs have ellipsoidal structure, and tend to be red and apparently passively evolving with limited star formation \citep[e.g.,][]{Visvanathan1977,bower+92}.  
While ETGs as a population appeared to be passive, many studies found signs of remnant cold gas reservoirs, and dust, in a subset of ETGs \citep{knapp+85,knapp+89,sage+89,welch+sage03}.  
Results from the $\atd$ survey, a volume-limited survey of nearby ETGs \citep{Cappellari2011}, indicate that ETGs {\em as a population} can contain large reservoirs of gas, including \hi, CO, HCN, and HCO$^+$  \citep{Young2011,Crocker2012,Serra2012}.  
The complete CO survey finds that at least $\sim 20$\% of ETGs have M(H$_2$) $> 10^7$~M$\odot$, with molecular gas reservoirs as large as $\sim 10^9$~M$_\odot$ in some ETGs \citep{Young2011}.
\begin{table*}
    \footnotesize
    \begin{center}
    \caption{Properties of Observed Galaxies} \label{tab:sources}
    \begin{tabular}{l r r r r r r r r r r}
        \hline
        \hline
        {\bf Name} & $\alpha$ & $\delta$ & $v_{\rm sys}$ & I$_{\rm CO}$ & I$_{\rm HCN}$ & $\lhcn / \lco$ & F$_{1.4}$ &  $\lfir$ & $S_{\rm 25\;\mu m}$ & $S_{\rm 100\;\mu m}$ \\
                   & (J2000) & (J2000) & (km s$^{-1}$) & (K km s$^{-1}$) & (K km s$^{-1}$) & & (mJy) & ($10^9 L_\odot$) & (Jy) & (Jy) \\
        \hline
        NGC 1266 & 03 16 00.8 & -02 25 38.4 & 2160 & 34.8 & 2.8 & 0.13 & 115.6 &  18 & 1.16 & 16.9 \\
        NGC 3665 & 11 24 43.6 & +38 45 46.1 & 2040 & 12.0 & 0.49 & 0.07 & 112.2 & 5.3 & 0.16 & 7.5\\
        NGC 5866 & 15 06 29.6 & +55 45 47.9 & 760 & 21.6 & 0.84 & 0.06 & 21.8 & 2.7 & 0.21 & 16.9 \\
        
         & & & & & & & \\
        NGC 2764 & 09 08 17.4 & +21 26 36 & 2707 & 16.2 & 0.28 & 0.03 & 28.1 & 10 & 0.48 & 7.2 \\
        NGC 3032 & 09 52 08.1 & +29 14 10 & 1549 & 8.3 & 0.27 & 0.05 & 7.2 & 1.8 & 0.19 & 4.7 \\
        IC 676 & 11 12 39.8 & +09 03 21 & 1429 & 11.6 & 0.27 & 0.04 & 9.7 & 3.1 & 0.81 & 4.9 \\
        NGC 3607 & 11 16 54.6 & +18 03 06 & 935 & 10.4 & 0.73 & 0.11 & 7.2  & 1.0$^a$& -- & -- \\
        NGC 4459 & 12 29 00.0 & +13 58 42 & 1169 & 10 & 0.59 & 0.10 & 1.8$^b$ & 1.0 & 0.30 & 4.8 \\
        NGC 4526 & 12 34 03.0 & +07 41 57 & 697 & 22 & 1.75 & 0.13 & 12.0 &  3.4 & 0.43 & 17.1 \\
        NGC 4710 & 12 49 38.8 & +15 09 56 & 1133 & 32 & 1.41 & 0.07 & 18.7 & 3.2 & 0.46 & 14.8 \\
        UGC 09519 & 14 46 21.1 & +34 22 14 & 1665 & 12.6 & 0.29$^c$ & $<$ 0.03 & 0.4$^{d}$ & 0.6 & $<$0.10 & 0.94 \\
        NGC 7465 & 23 02 01.0 & +15 57 53 & 1974 & 11.9 & 0.13 & 0.015 & 19.1 & 7.6 & 0.49 & 8.2 \\
        \hline \hline
    \end{tabular}
    \end{center}
    \vskip 0.25em
    \raggedright
    {\bf Note:} 
    At the top, the three galaxies observed with the GBT. NGC~1266 is the only $\atd$ galaxy previously observed in OH.  
    Below, the nine ETGs observed with Arecibo.  
    Eleven of the galaxies were detected in HCN, while UGC~09519 has an upper limit in HCN, and was detected in HCO$^+$. 
    CO fluxes are from \citet{Young2011} and HCN fluxes are from \citet{Crocker2012}. 
    Except where noted, infrared fluxes are from {\em IRAS}, and $\lfir$ is calculated according to \citet{Sanders1996}. 
    Integrated 1.4~GHz flux densities come from NVSS \citep{Condon1998}, except where noted; all radio structure is unresolved for the resolution of NVSS. 

$^a$ Estimated from {\em AKARI} fluxes, as it was not observed by {\em IRAS}. \\
$^b$ From \citet{Becker1995}. This source was included in our survey due to mistakenly transcribing its radio flux density. \\
$^c$ Measured fluxes are for HCO$^+$ rather than HCN. \\
$^d$ This source was included in the survey because we used the 2.2~GHz flux, 4.2~mJy, from \citet{Dressel1978}. More recent observations by Nyland et al. (2014, in prep.) find the lower flux density given here. \end{table*}

Though a small number of ETGs have been observed in OH, there has been no systematic search for OH in ETGs.
Observations of OH in ETGs provide an opportunity to better understand the excitation of OH, and test whether galaxies with high dense gas fractions can produce (weak) maser emission despite low FIR luminosities.
Studying OH in ETGs also has potential utility in understanding how star formation is quenched.  
Outflows of molecular gas are a promising means of removing star forming material from galaxies and producing quenched systems, and there are indications that OH is a promising tracer of outflows in ETGs. 
Previous work has used OH as an outflow tracer, in both the ground state 18~cm OH lines \citep{Baan1989a} and FIR OH lines \citep{Sturm2011}.
While galaxies with OH outflow detections are primarily LIRGs, there are two ETGs that show evidence for outflows in CO (NGC~1266, \citealt{Alatalo2011}; and NGC~1377, \citealt{aalto+12}), a detection of OH in absorption in NGC~1266 at 18~cm with $\tau \sim 0.1$ by \citet{Baan1992} shows hints of broad wings in OH.  

We aim to study the OH content and excitation in a population of galaxies in which there has been no systematic search for OH, while also using OH to search for evidence of outflows.
We motivate our selection of sources in Section \ref{sec:sources}, and then describe the observations in Section \ref{sec:data}.   We present our results in Section \ref{sec:results}, including three new detections of OH in absorption.   In Section \ref{sec:discussion}, we discuss how our detections and non-detections inform previous attempts to understand the presence and excitation of OH in galaxies.   Finally, we briefly summarize our main results in Section \ref{sec:conclusions}.

\section{Source selection} \label{sec:sources}
We selected sources from the $\atd$ project, which is a multi-wavelength survey of ETGs, defined to be types S0 and earlier \citep{Cappellari2011}.  
The work of \citet{Darling2007} suggests that the $\lfir$ and $\lhcn / \lco$ phase space is an important one to consider in understanding OH excitation in galaxies, so we focused on the subset of galaxies with detections in both CO(1-0) and HCN.
$\lhcn / \lco$ is used as a dense gas tracer because HCN is excited at higher gas densities than CO(1-0).  
\citet{Gao2004} found that typical star forming galaxies fall in a small range of dense gas fractions, with $\lhcn / \lco = 0.02$--0.05, while (U)LIRGs have higher dense gas fractions, with $\lhcn / \lco > 0.06$.  
The ETGs that \citet{Crocker2012} studied mostly had $\lhcn / \lco$ ratios in line with typical star forming galaxies, but a few in the sample had ratios more typical of (U)LIRGs.  
\citet{Crocker2012} suggested that high $\lhcn / \lco$ in ETGs may reflect optical depth effects, but Alatalo et al. (2014, submitted) find that $\lhcn / \lco$ likely still traces dense gas fraction in ETGs.

The large $\lfir$ values of galaxies with OHMs distinguish them from the ETGs observed as part of the $\atd$ survey. 
With low $\lfir$ and cool dust temperatures, ETGs are unlikely to be able to radiatively pump the OH molecule \citep{Lockett2008}.
Nonetheless, the most dense gas rich galaxies in the $\atd$ survey are in a poorly explored area of the $\lfir$ and $\lhcn / \lco$ phase space, providing a test for understanding OH excitation.
The $\atd$ galaxy with the highest dense gas fraction, NGC~1266, also shows evidence for an outflow, further motivating observing OH in galaxies with large $\lhcn / \lco$. 
Thus we selected twelve galaxies from the $\atd$ survey with detections of CO(1-0) \citep{Young2011} and HCN(1-0) \citep{Crocker2012}.\footnote{One galaxy, UGC 09519, was detected in HCO$^+$, but not in HCN.}
We also chose galaxies with large enough radio continuum fluxes to detect absorption $\tau \sim 0.1$ with velocity resolution $10\kms$ in a $\sim$$2$ hour observation.
All twelve sources are listed in Table \ref{tab:sources}, along with the gas properties measured in the $\atd$ survey.

\section{Observations and Data Reduction} \label{sec:data}
We used the $L$-band wide receiver at Arecibo Observatory\footnote{The Arecibo Observatory is operated by SRI International under a cooperative agreement with the National Science Foundation (AST-1100968), and in alliance with Ana G. M\'{e}ndez-Universidad Metropolitana, and the Universities Space Research Association.} to observe nine ETGs at declinations -1$^\circ$20' $< \delta < $ +38$^\circ$02'.
For three sources outside of this declination range, we used the Green Bank Telescope (GBT),\footnote{The Green Bank Telescope is operated by the National Radio Astronomy Observatory, which is a facility of the National Science Foundation operated under cooperative agreement by Associated Universities, Inc.} and selected the L-band receiver and the GBT Spectrometer in spectral line mode.
For both telescopes, we observed in Full-Stokes mode. 
Though we did not expect to detect polarized emission, full-Stokes data are often useful for radio frequency interference (RFI) identification.

The ground state of OH has four lines at wavelengths $\sim$18~cm as a result of $\Lambda$-doubling and hyperfine splitting. 
There are two ``main'' lines at 1665~MHz and 1667~MHz, and two ``satellite'' lines at 1612~MHz and 1720~MHz.  For OH in thermal equilibrium, the expected ratio of line strengths for the 1612:1665:1667:1720 MHz lines is 1:5:9:1. 
Main lines are typically the target of 18~cm OH observations because of their strength, but the spectrometers for both Arecibo and the GBT can be configured to capture four sub-bands within L-band.  
We set the sub-bands to observe the \hi\ line, both OH main lines, the 1612~MHz OH line, and the 1720~MHz OH line.  
For Arecibo, each sub-band was 12.5~MHz wide with 2048 channels, which provides a velocity resolution of $\sim 1 \kms$; the GBT sub-bands were 12.5~MHz wide with 4096 channels, corresponding to a $\sim 0.5 \kms$ velocity resolution.

\begin{table}
  \begin{center}
  \caption{Optical Depth Sensitivity} \label{tab:rms}
  \begin{tabular}{l r r r}
  \hline
  \hline
            & & rms / continuum \\ \hline
  NGC 3665  & & 0.09  & \\
  NGC 5866  & & 0.045  & \\   
            & & & \\
  IC 676    & & 0.037 & \\ 
  NGC 2764  & & 0.017 & \\
  NGC 3032  & & 0.037 & \\
  NGC 3607  & & 0.044 & \\
  NGC 4459  & & 0.06  & \\
  NGC 4526  & & 0.025 & \\
  NGC 4710  & & 0.020 & \\
  NGC 7465  & & 0.016 & \\
  UGC 09519 & & 0.17  & \\ \hline \hline
  \end{tabular}
  \end{center}
  \vskip 0.25em
  {{\bf Note:} The values for the ratio of the rms noise to the continuum flux correspond to channels with widths $\sim$10~$\kms$.
We can detect smaller optical depths than the values in the table; this metric simply compares the relative sensitivity to absorption sources with differing continuum fluxes, integration times, and telescopes.
The top two sources were observed with the GBT; the remainder were observed with Arecibo.
NGC~1266 is omitted from the table, due to severe non-astrophysical structure in the spectrum.
}
\end{table}
The sources are all unresolved for the 3' beam of Arecibo and the 8' beam of the GBT.  
We observed each source for $\sim$$2$ hours, split equally on and off source. 
We position switched at intervals of $4$ minutes for Arecibo and $2$ minutes for the GBT, using off source positions that were 0$^h$04$^m$ in right ascension behind the source positions.  
For both, we took integration times of $1$~s, to allow editing of short time duration RFI.  
We visually inspected all data in frequency/time space, flagging integrations with significant noise or structure, or with strong polarized emission.
We then calibrated the data using the RHSTK software package (Heiles et al., in prep.).
We adopted gains of 10~K~Jy$^{-1}$ for Arecibo\footnote{See \href{http://www.naic.edu/~astro/RXstatus/Lwide/Lwide.shtml}{http://www.naic.edu/\textasciitilde astro/RXstatus/Lwide/Lwide.shtml}} and 2~K~Jy$^{-1}$ for the GBT\footnote{See listing for Rcvr1\_2 at \href{http://www.gb.nrao.edu/~fghigo/gbtdoc/sens.html}{http://www.gb.nrao.edu/\textasciitilde fghigo/gbtdoc/sens.html}} to convert system temperatures to flux densities.

As a measure of the sensitivity of our observations, Table \ref{tab:rms} shows the ratio of the rms noise in channels with widths $\Delta v \sim$10~$\kms$ to the radio continuum for each of the observed sources.
We chose $\sim$$10 \kms$ because up to that resolution, noise is roughly Gaussian (binning channels reduces the noise as $\sqrt{\Delta v}$), whereas there can be minor bandpass structure for larger widths.  
We can detect absorption with smaller average optical depth over the line than the ratio of rms to continuum flux, as absorption features have widths $\sim$$200 \kms$.

\section{Results} \label{sec:results}
We detected OH absorption for the first time in NGC~4526, NGC~4710, and NGC~5866, and re-detected absorption in NGC~1266. 
The properties of the detections are presented in Table \ref{tab:dets}.
All detections are in the main lines; we did not detect satellite lines in any source.
For each absorption line of newly detected sources, we use bootstrap resampling to estimate the error in the measured optical depth.
The ``measurements'' from which we generate bootstrap samples are individual on/off scans (sampled every 4 minutes for Arecibo and every 2 minutes for the GBT; see \citealt{McBride2013} for more detail).
For each Arecibo source, we had $\sim$20 ``measurements'' ($\sim$40 for the GBT), which we then re-sampled 1000 times.
While one of our motivations was to search for outflows in OH, we did not have the sensitivity to detect any evidence of outflows.
In the following sub-sections, we discuss each detection individually, and present spectra of the new detections.

\subsection{NGC 1266}
NGC~1266 is a nearby FIR-bright ETG that appeared to be a passive S0 galaxy, until \citet{Alatalo2011} showed that NGC~1266 contained a mass-loaded molecular outflow and a dense molecular gas core. 
It also shows signs of star formation suppression as a result of AGN activity \citep{alatalo+14,Alatalo2014a}. 

\citet{Baan1992} detected OH absorption with asymmetric structure extending to the red for both OH main lines in NGC~1266.
Some sources similar to NGC~1266 in FIR color and luminosity, and with smaller dense gas fractions, have narrow maser emission in their absorption profiles.  
We re-observed NGC~1266 hoping to improve upon the SNR of the original detection and potentially see evidence for masers or OH in outflow.
Due to instrumentation issues, our data were of much poorer quality than the original spectrum. 
We use the literature data for subsequent discussion.

\subsection{NGC 4526}
\begin{figure}
    \includegraphics[width=3.4in]{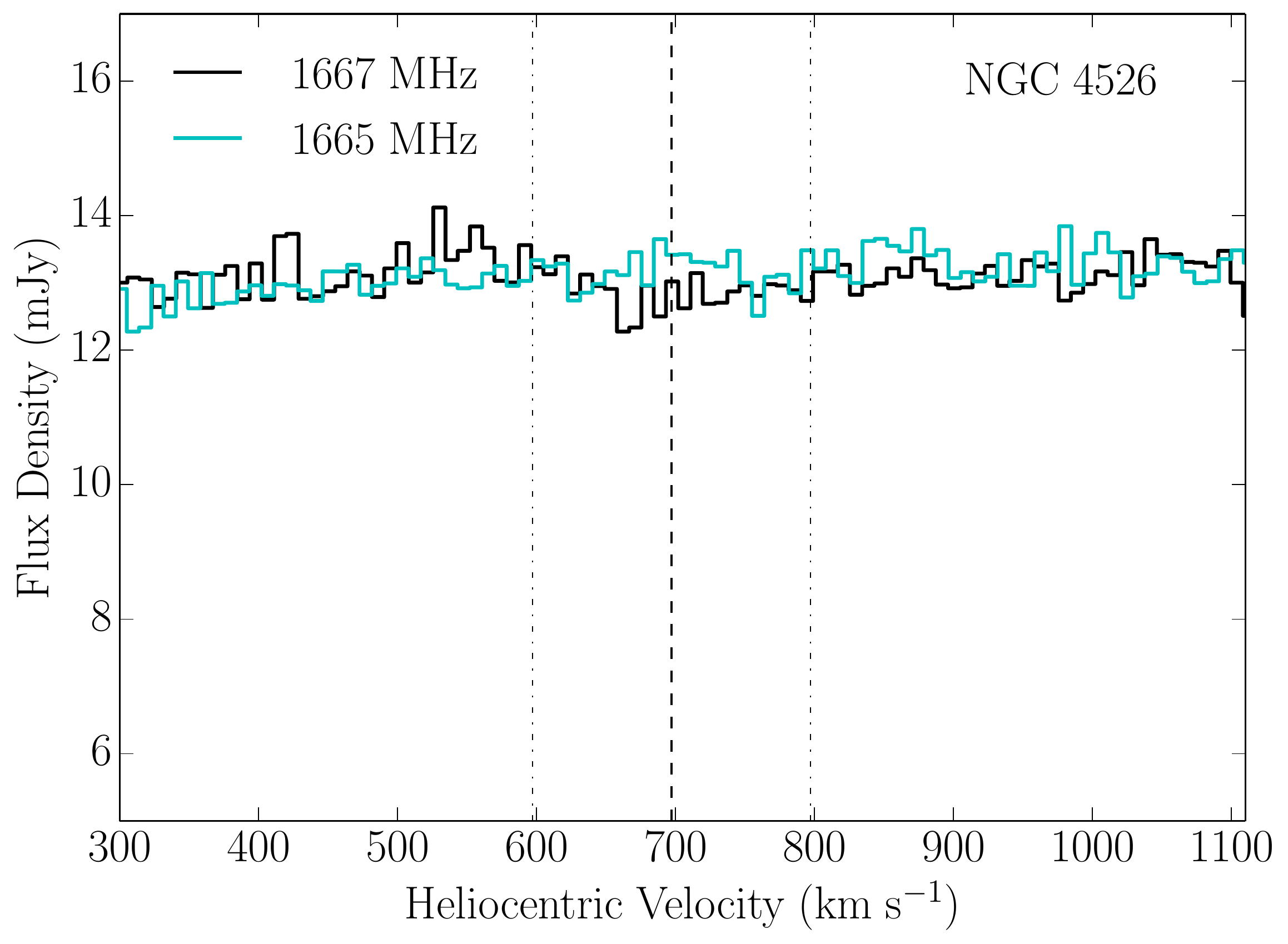}
    \caption{Profile of NGC 4526. Here, and in Figures \ref{fig:ngc4710} and \ref{fig:ngc5866}, the 1667~MHz (solid black) and 1665~MHz (solid cyan) lines are shown. A dashed black line shows the heliocentric velocity of the galaxy. The bounds of 1667~MHz absorption, identified by eye, are marked with thin black dash-dot lines. For NGC~4526, the spectra are binned into channels with widths of $8.9 \kms$.
} \label{fig:ngc4526}
\end{figure}

\begin{table*}
    \begin{center}
    \caption{OH Properties of Detected Galaxies} \label{tab:dets}
    \begin{tabular}{l r r r r r} \hline \hline
         & $v_{\rm sys}$ & $v_{\rm abs}$ & $\thi$ & $\tlo$ & N(OH) / $T_{\rm ex}$ \\ 
        & (${\rm km\;s}^{-1}$) & (${\rm km\;s}^{-1}$) & & & ($10^{15}$ cm$^{-2}$ / K) \\ \hline
        NGC 1266 & 2160 & 2000--2400 & 0.07 & 0.04 & 7.2 \\
        NGC 4526 & 697 & 597--797 & 0.026 $\pm$ 0.008 & 0.003 $\pm$ 0.009 & 1.3 $\pm$ 0.4\\
        NGC 4710 & 1133 & 1053--1213 & 0.033 $\pm$ 0.006 & 0.004 $\pm$ 0.007 & 1.3 $\pm$ 0.2 \\
        NGC 5866 & 760 & 720--840 & 0.08 $\pm$ 0.01 & 0.04 $\pm$ 0.01 & 2.3 $\pm$ 0.3 \\ \hline \hline
    \end{tabular}
    \end{center}
    \vskip 0.25em
    \raggedright

    {\bf Note:} The velocity range of absorption, $v_{\rm abs}$, is determined by eye, and the ranges in this table corresponded to the region between dash-dot lines in Figures \ref{fig:ngc4526}, \ref{fig:ngc4710}, and \ref{fig:ngc5866}. 
    $\bar{\tau}$ is the average optical depth within the given velocity range.
    We do not use bootstrap resampling to calculate error estimates for NGC~1266 for two reasons: there were only a few usable scans, and we can only distinguish absorption from non-astrophysical structure in the spectrum by comparison with the results of \citet{Baan1992}.
    The values of $\tau$ we measure for NGC~1266 are consistent with \citet{Baan1992}.
\end{table*}

NGC~4526 is a member of the Virgo cluster. 
It was originally imaged by \citet{Young2008}, who found CO(1--0) emission that appeared to be in a disk configuration, although the unsharp-masked Hubble Space Telescope image showed dust that appeared to be a ring.  
Higher resolution CO(2-1) observations supported evidence for the ring, and also found resolved giant molecular clouds near the center of the galaxy \citep{davis+13}.

We marginally detect absorption in the 1667~MHz OH line in NGC~4526.  
The spectra of both main lines are shown in Figure \ref{fig:ngc4526}. 
For the 1667~MHz line, we measured an average optical depth $\thi = 0.026 \pm 0.008$ between the region marked by black dash-dot lines.
We do not detect the 1665~MHz line.
Though this detection is not visually convincing, two pieces of evidence suggest weak absorption.
We randomly selected other velocity ranges in the spectrum of NGC~4526 over which to calculate an optical depth of absorption and estimate errors; we did not find other features with non-zero $\tau$.
We also convolved the spectrum with tophat filters with widths 100,~125,~150,~...,~400~$\kms$.
We calculated a $\tau$ at each lag using the continuum flux across the whole spectrum and the flux within the tophat. This procedure produced a peak in $\tau$ at a velocity within a few $\kms$ of the velocity of NGC~4526 for all tophat widths $< 250 \kms$.
For these tophat widths, the $\tau$ at the velocity of the galaxy was a factor of $\sim$3 larger than $\tau$ at any other local maximum in the convolved spectrum.

\subsection{NGC 4710}

\begin{figure}
    \includegraphics[width=3.4in]{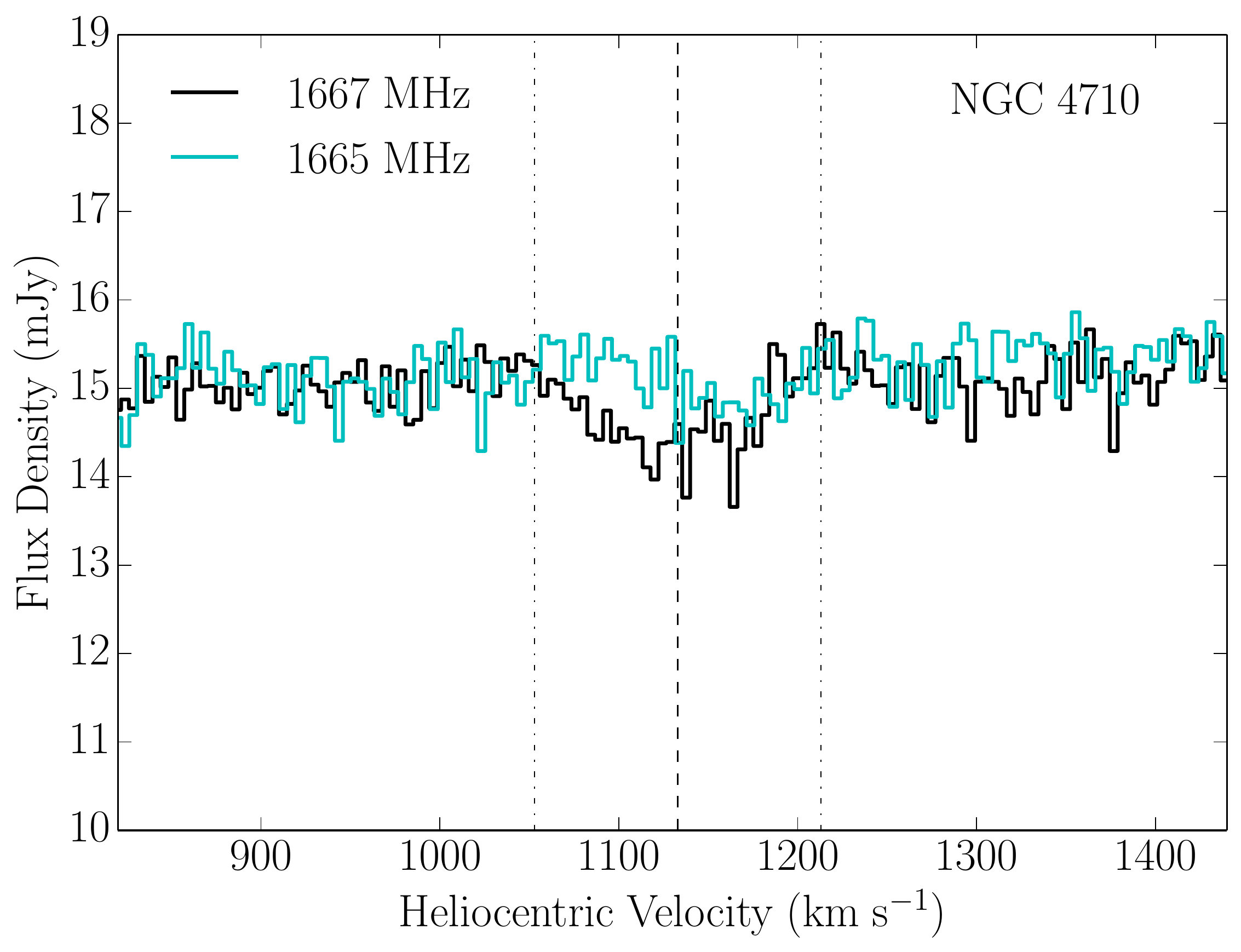}
    \caption{Profile of NGC 4710.
Labels are the same as in Figure \ref{fig:ngc4526}. The OH spectra are binned into channels with widths of $4.5 \kms$. 
} \label{fig:ngc4710}
\end{figure}

NGC~4710 is an edge-on galaxy at the outskirts of the Virgo cluster, and has a prominent molecular gas bar \citep{alatalo+13}.
The spectra for the OH main lines are shown in Figure \ref{fig:ngc4710}.  
Within the dash-dot lines, we find an average optical depth $\thi = 0.033 \pm 0.006$ for the 1667~MHz line. 
The 1667~MHz line has roughly symmetric structure, with equal absorption (within error) above and below the system velocity.
The average optical depth for the 1665~MHz line is consistent with a non-detection.
However, there is tentative evidence that the 1665~MHz line absorbs more strongly at velocities redshifted relative to the velocity of the galaxy galaxy.  
For $v = 1133$--$1213 \kms$, $\tlo = 0.015 \pm 0.007$ and $\thi = 0.030 \pm 0.009$, giving a 1667/1665 ratio $2 \pm 1$.
For $v = 1053$--$1133 \kms$, $\tlo = -0.01 \pm 0.01$ and $\thi = 0.035 \pm 0.009$.

For OH in local thermodynamic equilibrium (LTE), the 1667/1665 ratio ranges from 1--1.8 depending upon optical depth.
Diffuse OH that is not in LTE can produce a 1667/1665 absorption ratio that exceeds 1.8.
There are some examples of 1667/1665 ratios moderately larger than 1.8 in other galaxies \citep{Baan1985a,Schmelz1986}, and larger departures from LTE have been observed in diffuse clouds in the Milky Way \citep{Dickey1981}.
With large uncertainties, we cannot definitively identify OH in different phases, but the observed spectrum is consistent with absorption from $v = 1133$--$1213 \kms$ from a denser OH phase in LTE, and an OH phase at $v = 1053$--$1133 \kms$ that is not in LTE.

\subsection{NGC 5866}
\begin{figure}
    \includegraphics[width=3.4in]{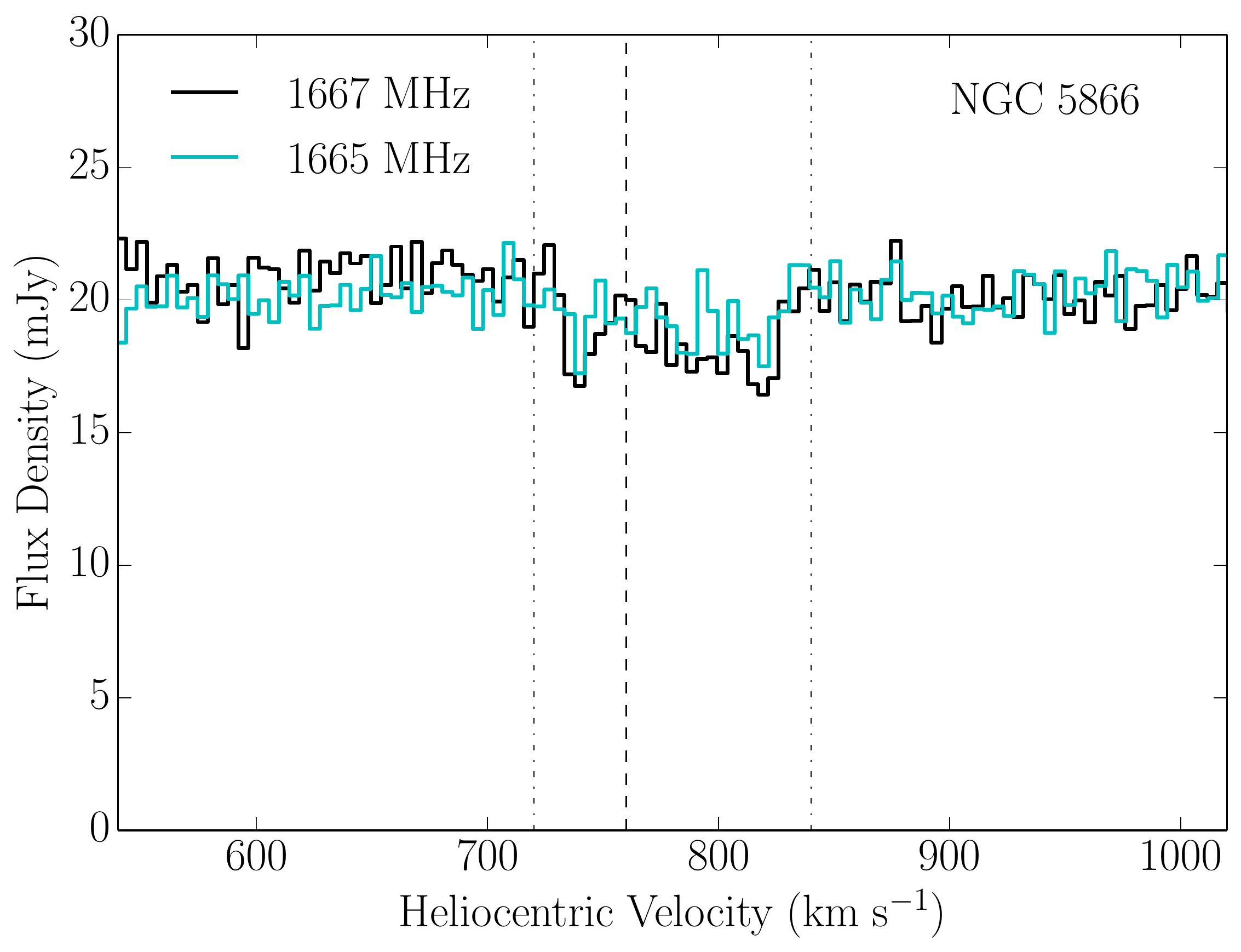}
    \caption{Profile of NGC 5866. 
Labels are the same as in Figure \ref{fig:ngc4526}. The OH spectra are binned into channels with widths of $4.5 \kms$.
} \label{fig:ngc5866}
\end{figure}

NGC~5866 is a nearby edge-on S0 field galaxy that has a strong bar in molecular gas \citep{alatalo+13}. 
Figure \ref{fig:ngc5866} shows the OH main lines for NGC~5866.  
Of the three new detections, NGC~5866 is the highest SNR, and the only source for which the structure in both the 1665~MHz and 1667~MHz lines is visibly similar.
The average optical depths in the region between the dash-dot lines in Figure \ref{fig:ngc5866} are $\thi = 0.08 \pm 0.01$ and $\tlo = 0.04 \pm 0.01$.
The 1667/1665 ratio is $2.0 \pm 0.5$ is consistent with absorption by optically thin OH, though the uncertainty is too large to rule out absorption by optically thick OH.
The optical depth appears to be double peaked, and asymmetric, with stronger absorption from the component redshifted with respect to the velocity of the galaxy.  
This is consistent with the presence of two velocity structures in the gas along the line-of-sight and the observed bar morphology.
The bar drives gas to inflow to the nucleus, producing the redshifted component, while absorption in the blueshifted component occurs in the disk \citep{alatalo+13}. 

\section{Discussion} 
\label{sec:discussion}
We detected OH in absorption for the first time in three galaxies: NGC~4526, NGC~4710, and NGC~5866.  
The sources with new detections generally have good sensitivity to absorption, but sensitivity alone does not account for the detections; multiple sources with comparable absorption sensitivity (i.e., IC~676, NGC~2764, NGC~3032, NGC~7465) are non-detections.  
The overall detection rate, 3/11, compares favorably with past surveys of OH \citep{Baan1992,Darling2002a}, though our sensitivity was also better.
The ETGs that we observed have values of $\lfir$, $\lco$, and $\lhcn$ that are unlike any previously observed sample, and suggests that it is feasible to discover OH in a wider range of galaxies than previous work indicated. In particular, the only previously detected OH absorber with smaller $\lir$ than the three new OH absorption detections is NGC~5363.

\subsection{The role of FIR luminosity and dense gas fraction in detecting OH}
\begin{figure}
\includegraphics[width=3.4in]{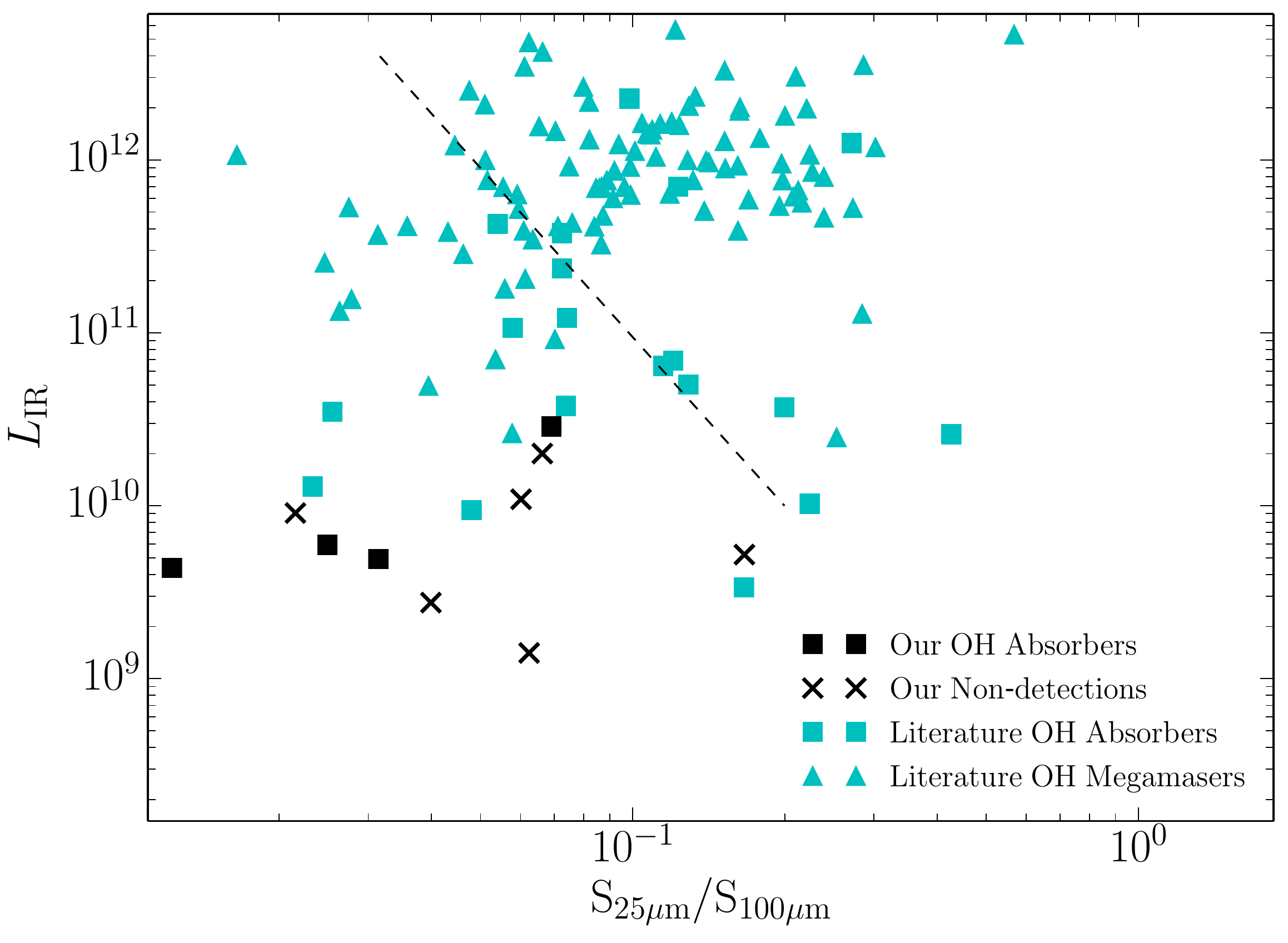}
\caption{Update of Figure 2 from \citet{Baan1989}, with our new observations added, as well as additional observations from the literature since his work. The ratio of the 25 micron flux to the 100 micron flux, $\log(S_{\rm 25\;\mu m} / S_{\rm 100\;\mu m})$, is an indication of the dust temperature, with larger values for warmer dust. The dashed line marks the separation between OHMs and absorbers that \citet{Baan1989} identified.
Non-detections of OH from the literature are not shown, as they greatly outnumber detections of absorption or maser emission, and make the figure more difficult to read.  
} \label{fig:ir_colors}
\end{figure}

In examining the excitation of OH in galaxies, \citet{Baan1989} looked at the dust temperatures and IR luminosities of OH absorbers and masers. 
In Figure \ref{fig:ir_colors} here, we update Figure 2 from \citet{Baan1989}.
All of our sources fall below and to the left of the dashed line that \citet{Baan1989} drew separating masers and absorbers.  
Along with low $\lir$ relative to previously detected OH absorbers, our galaxies also have very cool dust temperatures.\footnote{\citet{Baan1989} used IR/FIR interchangeably, but appear to have calculated $\lir$ from all four {\em IRAS} bands. We follow suit, adopting the $\lir$ definition from \citet{Sanders1996}. Using $\lfir$ instead would not affect the result.} 
Thus our detection of absorption, rather than maser emission, agrees with the conclusions of \citet{Baan1989}. 

\begin{figure}
\includegraphics[width=3.4in]{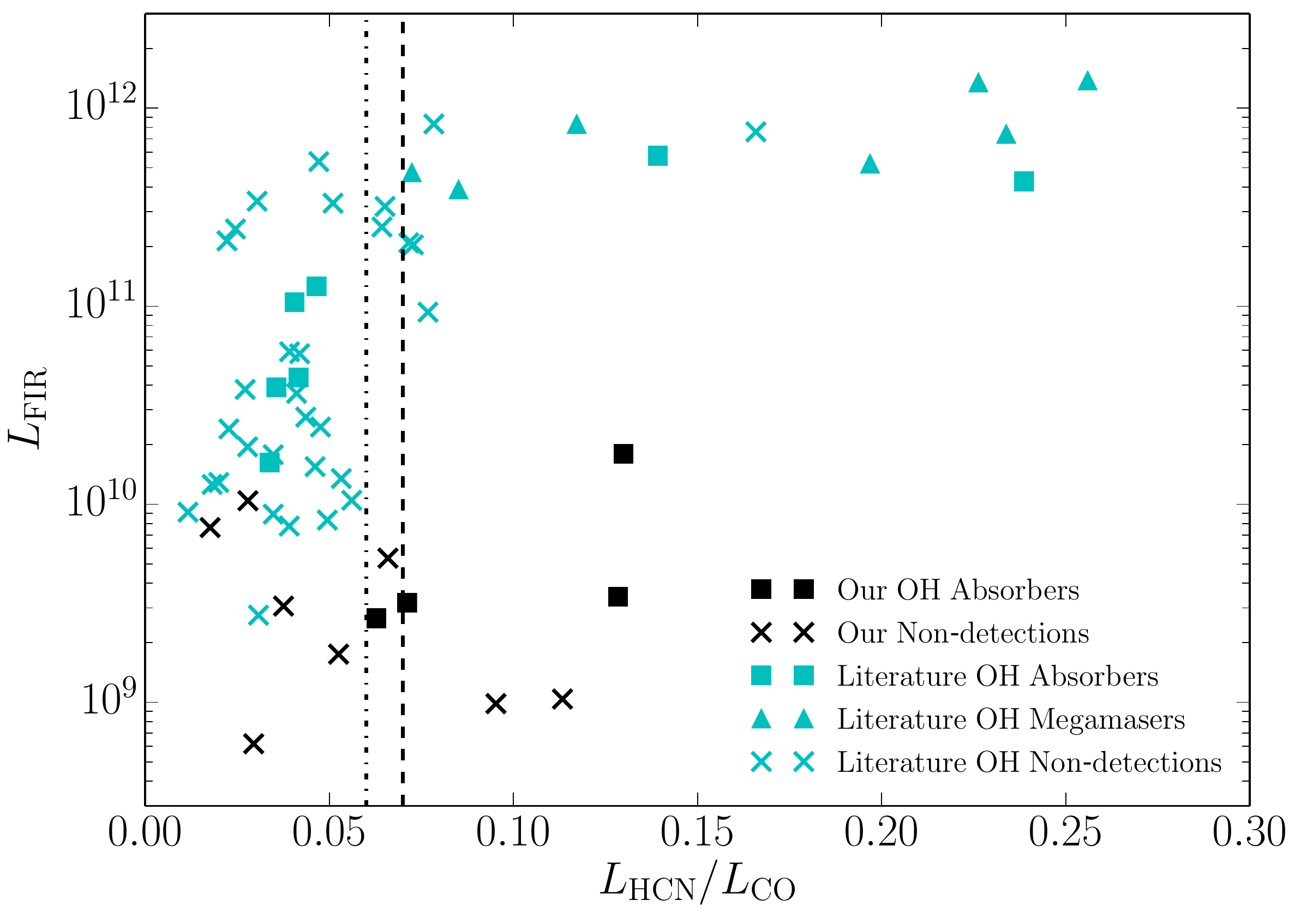}
\caption{Update of Figure 2 from \citet{Darling2007}, with our new observations added. 
The target galaxies fall in an unexplored part of this phase space, at lower $\lfir$, but with moderate to large $\lhcn / \lco$.
The black dashed line at $\lhcn / \lco = 0.07$ shows the threshold that \citet{Darling2007} identified for producing OHMs.
The dash-dotted line at $\lhcn / \lco = 0.06$ shows the separation between normal star forming galaxies and LIRGs that \citet{Gao2004} identified. 
} \label{fig:hcn-co}
\end{figure}

\citet{Darling2007} argued that high dense gas fraction drives OHM activity more than large $\lfir$, as half of galaxies with $\lhcn / \lco > 0.07$ hosted OHMs.
Our sample included four ETGs above that threshold, and in three of these we detected absorption.
Along with two other previously known OH absorbers, there are now five sources with $\lhcn / \lco >$ 0.07 that absorb OH.
These are shown in Figure \ref{fig:hcn-co} (an updated version of Figure 2 from \citealt{Darling2007}).
As even ETGs with large $\lhcn / \lco$ have much smaller gas reservoirs than LIRGs, the physical conditions of our OH detections are different from typical LIRGs. 
The ETGs in our sample mostly fall below the $\lfir$-$\lhcn$ relation in \citet{Gao2004}. 
We can apply the model relating $\lfir$ and $\lco$ from \citet{Krumholz2007} to estimate mean H$_2$ densities in ETGs and OHM hosts. 
This procedure suggests that the ETGs with large $\lhcn / \lco$ have mean H$_2$ densities somewhat lower than most OHM hosts with similar values of $\lhcn / \lco$, and in a smaller volume of gas. 
Nevertheless, the mean H$_2$ densities in the ETGs are still comparable to the least dense of the OHM hosting LIRGs.
This indicates that detecting exclusively absorption in ETGs with high $\lhcn / \lco$ is a result of how much less FIR luminous ETGs are than the galaxies that \citet{Darling2007} considered. 
Dense gas can only trigger OHMs in systems with the high dust temperatures and FIR fluxes required to radiatively pump OH \citep{Lockett2008}.

The four detections of OH absorption in this survey occurred among the seven largest $\lhcn/\lco$ ratios, and with $\lhcn/\lco$ near or above the threshold identified by \citet{Darling2007}.
Compared to the rest of our sample, the detected galaxies had average $\lfir$ and relatively cool dust temperatures.
These results suggest that $\lhcn / \lco$ is a useful diagnostic for identifying systems with OH, either in absorption or emission, regardless of $\lfir$, but
dust temperature and $\lfir$ are important in determining whether OH will be seen in absorption or maser emission.  

\subsection{Comparison of observed galaxy properties}
NGC~1266 is much more akin to ``typical'' OH absorbers, with strong radio continuum \citep{nyland+13}, an AGN-driven molecular outflow, and the largest $\lfir$ in our sample. For these reasons, NGC~1266 is the least surprising OH detection.  The following discussion therefore focuses on the remaining ETGs in the sample.

Though absorption sensitivity improves with larger radio continuum fluxes, the varying sensitivity of our observations is not a significant determinant of whether a galaxy was detected or not. 
Table \ref{tab:rms} shows that detections and most non-detections have comparable ratios of rms error to continuum flux.
A favorable alignment between gas and radio continuum is more important than having a large radio continuum flux.
While many of our sources have radio jets, all but NGC~3665 are core dominated (Nyland et al. 2014, in prep.).
Thus centrally concentrated gas morphologies, and edge-on inclinations, are important for producing OH absorption.

\citet{alatalo+13} used CARMA to image molecular gas and discuss gas morphology for all the sources in our survey.
Two of the three detected galaxies (NGC~4710 and NGC~5866) have bar+ring morphologies. 
Though NGC~4526 has a disk morphology, \citet{davis+13} showed that it has a nuclear ring.
Among eight non-detections, only one, NGC~2764, is a bar+ring galaxy, and it also has a low $\lhcn/\lco$. 
All other non-detected galaxies were either disk morphologies or mildly disrupted (minor merger-like).  
Bar systems are capable of funneling gas into the center of galaxies, which could explain the presence of sufficient gas near the radio continuum emitting source to produce OH absorption. 
Using inclination angles reported in \citet{davis+11} (for NGC~5866, instead see \citealt{alatalo+13}), we find all three OH detections have $i>80^\circ$.
The non-detections had inclination angles ranging between $i=35-70^\circ$, much more face on than the detected objects.  
Thus all OH detections have molecular gas coincident with the radio continuum emission along the line-of-sight.

Altogether, OH absorption requires three interrelated elements: dense molecular gas along the line-of-sight to a radio continuum source. 
Favorable lines-of-sight can be produced by edge on inclinations (e.g., NGC~5866) or gas immediately surrounding the radio continuum source (e.g., NGC~1266). 
These conditions are satisfied in a wider range of galaxies than have been previously targeted in observations of OH. 

\subsection{OH abundances}
The abundance of OH relative to H$_2$ has not been well characterized in many astrophysical environments, and represents a significant uncertainty in converting measurements of OH to estimates of mass outflow rates in galaxies \citep[e.g.,][]{Sturm2011}.
Though the physical conditions in outflows likely differ from the conditions experienced by OH we detected, it is useful to expand the range of conditions for which estimates of OH abundances exist. 
We provide estimates of OH/H$_2$ ratios for galaxies with detected OH in Table \ref{tab:ratios}.

H$_2$ columns were derived from the CO(1--0) measurements of \citet{alatalo+13} along the line-of-sight to the radio continuum source from Nyland et al. (2014, in prep.).  
The brightness temperature in the integrated intensity map was taken in the pixel corresponding to the radio point source, and converted into N(H$_2$) using the X$_{\rm CO} = 2\times10^{20}$~cm$^2$~(K~km~s$^{-1})^{-1}$ conversion factor \citep{Bolatto2013}. 
We take only the central pixel corresponding to the radio point source because the OH absorption occurs against the nuclear radio continuum.
To calculate an OH column density, we assume an excitation temperature $T_{\rm ex} = 20$~K for all galaxies, based on previous work that has assumed values $T_{\rm ex} \sim 10$--$40$~K \citep[e.g.,][]{Baan1985a,Omar2002}.
We estimate 50\% uncertainty in the H$_2$ column densities, which is dominated by systematic uncertainty in X$_{\rm CO}$ and in absolute flux calibration.
For the OH column densities, we estimate a factor of two uncertainty, dominated by the assumed $T_{\rm ex}$. 
Altogether, we take a factor of three uncertainty in the ratio of OH/H$_2$.

\begin{table}
    \caption{OH/H$_2$ for detected galaxies} \label{tab:ratios}
    \centering
    \begin{tabular}{l r r r} \hline \hline
        & N(OH) & N(H$_2$) & N(OH) / N(H$_2$) \\ \hline
        & ($10^{16}$ cm$^{-2}$) & ($10^{22}$ cm$^{-2})$ & ($10^{-7}$)\\
      NGC 1266 & 14 & 390 & 0.4 \\
      NGC 4526 & 2.6 & 3.5 & 7 \\
      NGC 4710 & 2.6 & 8.6 & 3 \\
      NGC 5866 & 4.6 & 9.0 & 5 \\ \hline \hline 
    \end{tabular}
\end{table}

Our estimates of N(OH)/N(H$_2$) range from 0.4--7~$\times 10^{-7}$, with the OH abundance in NGC~1266 well below the abundance in the other detected galaxies. 
We attribute the discrepancy to either an unusual OH abundance or a significantly different excitation temperature than we assumed, as we consider an unusual value of $X_{\rm CO}$ in NGC~1266 unlikely \citep{Alatalo2011}.
If the excitation temperature of OH is higher in NGC~1266 than in the other galaxies, then we will have underestimated the OH abundance. 
Given evidence for shocks in NGC~1266 \citep{Pellegrini2013}, a higher excitation temperature than the 20~K we assumed is likely, though shocks can also enhance OH abundance \citep{Wardle1999}. 

Unfortunately, more detailed interpretation of these values is difficult. 
In addition to uncertainty in our abundance measurements, models of molecular abundances can produce similar predictions for OH abundances in different conditions. 
For instance, it is possible to produce N(OH)/N(H$_2$) $\sim 10^{-7}$ in models characteristic of dense clouds ($n_{H_2} \sim 10^4$--$10^6$ cm$^{-3}$) in the Milky Way \citep[e.g.,][]{Bergin1995,Sternberg1995}, or in models of more diffuse clouds \citep[e.g.,][]{vanDishoeck1986}, with $n_{H_2} \sim 10^3$.
Our measured abundances are thus consistent with gas in either diffuse or dense phases.

\section{Conclusions} \label{sec:conclusions}
We observed OH in early-type galaxies with significant reservoirs of molecular gas, as measured by HCN and CO.  This selection of galaxies produced three new detections of OH in absorption, out of a total of eleven sources that were observed in OH for the first time.  One goal of this study was discovering outflowing gas, but the SNR of the detected sources was inadequate to see evidence of outflows, such as broad absorption wings.
Our results do suggest an approach to future discovery of OH in ETGs in galaxies unlike those typically targeted for observations of OH. 
Including NGC~1266, which had previously been detected, our four absorption detections occurred among the seven galaxies with the highest ratios of $\lhcn / \lco$.  All of the sources in the sample, including the detections, are among the least FIR luminous sources with observations of OH.
Both large $\lfir$ and large $\lhcn / \lco$ appear necessary to produce large scale masing in galaxies.
A combination of dense gas, a radio continuum source, and a favorable line-of-sight (either through galaxy inclination or proximity of gas to the continuum source) improve the likelihood of detecting OH absorption. 

\section{Acknowledgments}
We are grateful to the referee for thoughtful comments that improved the content and clarity of the paper.
We thank Andrew Siemion for assistance observing remotely with the GBT, and Carl Heiles and Leo Blitz for useful discussions.
J. M. received support from a National Science Foundation Graduate Research Fellowship.  K. A. is supported by funding through {\em Herschel}, a European Space Agency Cornerstone Mission with significant participation by NASA, through an award issued by JPL/Caltech.
 This research used NASA Astrophysics Data System Bibliographic Services, the SIMBAD database, operated at CDS, Strasbourg, France, and the NASA/IPAC Extragalactic Database (NED), which is operated by the Jet Propulsion Laboratory, California Institute of Technology, under contract with NASA.

\bibliographystyle{mn2ealt}
\bibliography{ms}

\appendix
\section{Detection of \hi\ from a high velocity cloud in the NGC~4526 spectrum}
\begin{figure}
    \includegraphics[width=3.4in]{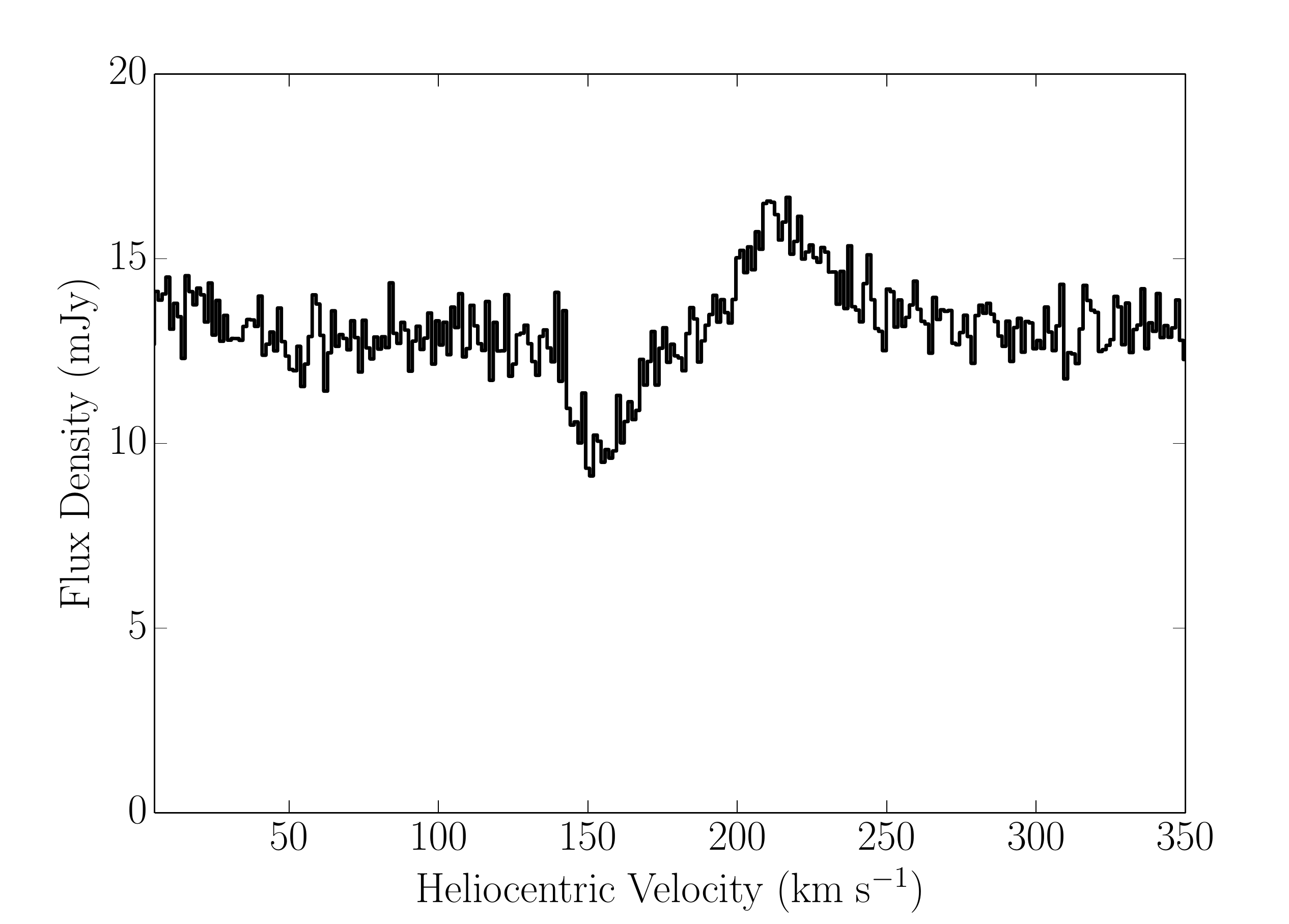}
    \caption{The \hi\ spectrum for NGC~4526 includes the apparent absorption/emission profile seen here. 
    We interpret the apparent absorption as emission in the position-switched off spectrum, which is then subtracted in generating the calibrated spectrum.
} \label{fig:hvc}
\end{figure}
We detect \hi\ in the spectrum for NGC~4526, though not \hi\ associated with the galaxy itself.  
To our knowledge, the best upper limit on the \hi\ mass in NGC~4526 is from \citet{Lucero2013}.
They used the Very Large Array to place an upper limit of $1.9 \times 10^7 M_\odot$ on the \hi\ mass in NGC~4526.  
We follow the same procedure they used for estimating the \hi\ mass upper limit, and find $M_{\rm H\; I} < 9 \times 10^6 M_\odot$.

Though we did not detect \hi\ associated with NGC~4526, we do see an apparent absorption/emission feature centered at a velocity of $\sim$200~$\kms$, shown in Figure \ref{fig:hvc}.
In observations toward SN~1994D, for which NGC~4526 is the apparent host galaxy, \citet{King1995} observed Na and Ca in absorption from high velocity clouds (HVCs) over a velocity range $v = 204$--$254 \kms$.
\citet{Wakker2001} observed SN 1994D at 21~cm as part of a survey of HVCs, but did not have sufficient sensitivity to detect the cloud in \hi.  
The velocity range over which we observed emission corresponds reasonably well with the velocity where Na and Ca occurred, and thus we interpret our apparent absorption/emission as a result of the HVC appearing in both our on and off spectra, and velocity structure in the cloud. 
From our detection, we estimate a column density N(\hi) $\sim 2 \times 10^{18} \cms$, though the column density may be higher if weak emission in the off spectrum occurs at the velocity of emission in the on spectra.

\end{document}